\documentclass[superscriptaddress,twocolumn, preprintnumbers, nofootinbib,amsmath,amssymb]{revtex4}

\pdfoutput=1

\usepackage{graphicx}
\usepackage{dcolumn}
\usepackage{bm}

\newcommand\ee{\end{equation}}
\newcommand\be{\begin{equation}}
\newcommand\eea{\end{eqnarray}}
\newcommand\bea{\begin{eqnarray}}

\newcommand\mpl{M_{\rm Pl}}

\begin{document}


\title{ISO(4,1) Symmetry in the EFT of Inflation}

\author{Paolo Creminelli}
\affiliation{Abdus Salam International Centre for Theoretical Physics,\\ Strada Costiera 11, 34151, Trieste, Italy}
\author{Razieh Emami}
\affiliation{School of Physics, Institute for Research in Fundamental Sciences (IPM), \\
P.~O.~Box 19395-5531, Teheran, Iran}
\affiliation{Abdus Salam International Centre for Theoretical Physics,\\ Strada Costiera 11, 34151, Trieste, Italy}
\author{Marko Simonovi\'c}
\affiliation{SISSA, via Bonomea 265, 34136, Trieste, Italy}
\affiliation{Istituto Nazionale di Fisica Nucleare, Sezione di Trieste, I-34136, Trieste, Italy}
\author{Gabriele Trevisan}
\affiliation{SISSA, via Bonomea 265, 34136, Trieste, Italy}


\begin{abstract}
In DBI inflation the cubic action is a particular linear combination of the two, otherwise independent, cubic operators $\dot\pi^3$ and $\dot\pi (\partial_i\pi)^2$. We show that in the Effective Field Theory (EFT) of inflation this is a consequence of an approximate 5D Poincar\'e symmetry, ISO(4,1), non-linearly realized by the Goldstone $\pi$.
This symmetry uniquely fixes, at lowest order in derivatives, all correlation functions in terms of the speed of sound $c_s$. In the limit $c_s \to 1$, the ISO(4,1) symmetry reduces to the Galilean symmetry acting on $\pi$. On the other hand, we point out that the non-linear realization of SO(4,2), the isometry group of 5D AdS space, does not fix the cubic action in terms of $c_s$.
\end{abstract}

\maketitle


\noindent

{\em Motivations.}
The study of non-linearly realized symmetries in the context of inflation has proven to be a powerful tool to make model-independent predictions. A spontaneously broken symmetry is manifested in relations among operators with different number of fields: for example, in the framework of the EFT of inflation \cite{Cheung:2007st} one finds a relation between the kinetic term and the cubic operators, as a consequence of the non-linear realization of time diffeomorphisms. This implies that in any model with small speed of sound $c_s\ll1$, one has parametrically large non-Gaussianities $\propto c_s^{-2}$. This regime is still allowed by observations, although severely constrained by the beautiful Planck data  \cite{Ade:2013ydc}.

In this note we study the consequences of the non-linear realization of ISO(4,1), the 5D Poincar\'e symmetry, in the EFT of inflation. The motivation is twofold. On one hand this symmetry is typical of inflationary models based on brane constructions, where the position of a brane moving in an extra dimension plays the role of the inflaton. Although the inflationary solution (spontaneously) breaks ISO(4,1), the dynamics of perturbations is constrained by the non-linearly realized symmetries.  
On the other hand, observations are only sensitive to small perturbations around the inflating solution and their dynamics is encoded in the EFT of inflation. It is then of interest to study the possible symmetries that can be imposed in this theory. In this respect ISO(4,1) naturally stands out, since it contains both the 4D Poincar\'e group and the shift symmetry of the inflaton, which is usually imposed to justify slow-roll and the consequent approximate scale-invariance of the spectrum.  We will show, for example, that the relation between the cubic operators $\dot \pi^3$ and $\dot \pi (\partial_i\pi)^2$ which occurs in DBI inflation \cite{Alishahiha:2004eh} does not require any UV input, but it is just a consequence of the ISO(4,1) symmetry at the level of the EFT of inflation. 

\vspace{.3cm}

{\em Nonlinear realization of ISO(4,1).}
In general, the homogeneous inflaton background $\phi_0(t)$ breaks the 4D Poincar\'e symmetry to translations and rotations: ISO(3,1) $\to$ ISO(3). (We here concentrate on scales much shorter than the Hubble scale $H$, where spacetime can be considered flat; we will consider gravity later on.)  
At leading order in slow-roll, the inflaton $\phi$ is also endowed with an approximate shift symmetry $\phi \to \phi +c$ and a solution $\phi_0(t) = v t$  preserves a combination of this shift symmetry and time translations.  

Perturbations around this background can be parametrized by the Goldstone mode $\pi$
\be
\phi(\vec x, t) = \phi_0(t+ \pi(\vec x,t))  = v \cdot (t+\pi) 
\ee
and the most general action compatible with the symmetries reads
\begin{align}
\label{effective1}
S = \int & \mathrm d^4 x \left( a_0 \pi + a_1 \dot{\pi}^2+a_2 (\partial_i\pi)^2+f_1 \dot{\pi}^3 +f_2 \dot \pi(\partial_i\pi)^2 \right. \nonumber \\
& \left. + g_1 \dot\pi^4 + g_2\dot\pi^2(\partial_i\pi)^2 + g_3 (\partial_i\pi)^4 + \cdots \right)\ .
\end{align}
All the constants are time independent as a consequence of the residual shift symmetry\footnote{The observed deviation from exact scale-invariance \cite{Ade:2013uln} implies that the shift symmetry (and therefore the whole ISO(4,1) in the following) is not exact, but slightly broken by corrections of order slow-roll. We neglect these corrections in the paper.}. 

Let us now impose the extra symmetry. We want to enlarge ISO(3,1) $\times$ shift (11 generators) to a 15-dimensional group, ISO(4,1). The additional four transformations act as\footnote{Notice that we are using a parametrization where the 4D coordinates do not transform and the symmetry only acts on fields.}
\be
\delta \phi= \omega_\mu x^\mu +\phi \;\omega^\mu \partial_\mu \phi \;.
\ee
These are rotations and boosts in the 5th dimension, if we interpret $\phi$ as a coordinate in the extra dimension, for example describing the position of a brane. The shift symmetry of $\phi$ is interpreted as translation in the 5th dimension to complete the isometry group of 5D flat space. However, the geometric interpretation is not mandatory and we may remain agnostic about the origin of this symmetry.  
These transformations act on the Goldstone $\pi$ as\footnote{In general $\phi_0(t) = c + v t$, but because of the shift symmetry the constant can be set to zero without loss of generality.} 
\begin{equation}
\label{transf1}
\delta \pi=\frac{1}{v}\delta \phi= \omega_\mu x^\mu + v^2 \cdot (t+\pi) (\omega^\mu \partial_\mu \pi + \omega^0)  \;,
\end{equation}
where in the last equality we have reabsorbed $1/v$ into the definition of $\omega_\mu$.
Demanding that the action \eqref{effective1} is invariant under these additional transformations imposes some conditions on the coefficients $a_0, \;a_1,\;a_2,\ldots$(\footnote{Notice that the tadpole term $a_0$ will in general be different from zero, since the background solution will also be affected by terms which are not ISO(4,1) symmetric, as a potential term and the Hubble friction. Anyway $a_0$ does not enter in the conditions below since its variation, eq.~\eqref{transf1}, is a total derivative.}).
If we focus on the variation of the action quadratic in $\pi$, we get the following relations 
\begin{equation}
\label{cubicrel}
a_2=-a_1(1-v^2)\;, \quad f_1=a_1\frac{v^2}{1-v^2} \;, \quad f_2=-a_1v^2 \;.
\end{equation}
The first equation says that the speed of propagation of $\pi$ excitations, the ``speed of sound" $c_s$ is related to $v$ as
\be
\label{cs2v}
c_s^2 = 1 - v^2 \;.
\ee
From the 5D geometrical point of view, this is a consequence of the relativistic sum of velocities. Here it is simply a consequence of the ISO(4,1) symmetry in the EFT of inflation. The cubic action is fixed by the second and third relation, so that up to cubic order the action (up to an overall coefficient) reads
\be
S = \int \mathrm d^4 x \,\left(\dot\pi^2 - c_s^2 (\partial_i\pi)^2 + \frac{1-c_s^2}{c_s^2}\left( \dot{\pi}^3-c_s^2 \dot \pi(\partial_i\pi)^2\right) \right)\ .
\ee
This is exactly the same result one gets in DBI inflation \cite{Alishahiha:2004eh}, but here we see that one does not need any UV input: this action follows from the ISO(4,1) symmetry in the EFT of inflation.

\begin{figure}[!!!h]
\begin{center}
\includegraphics[width=0.45 \textwidth]{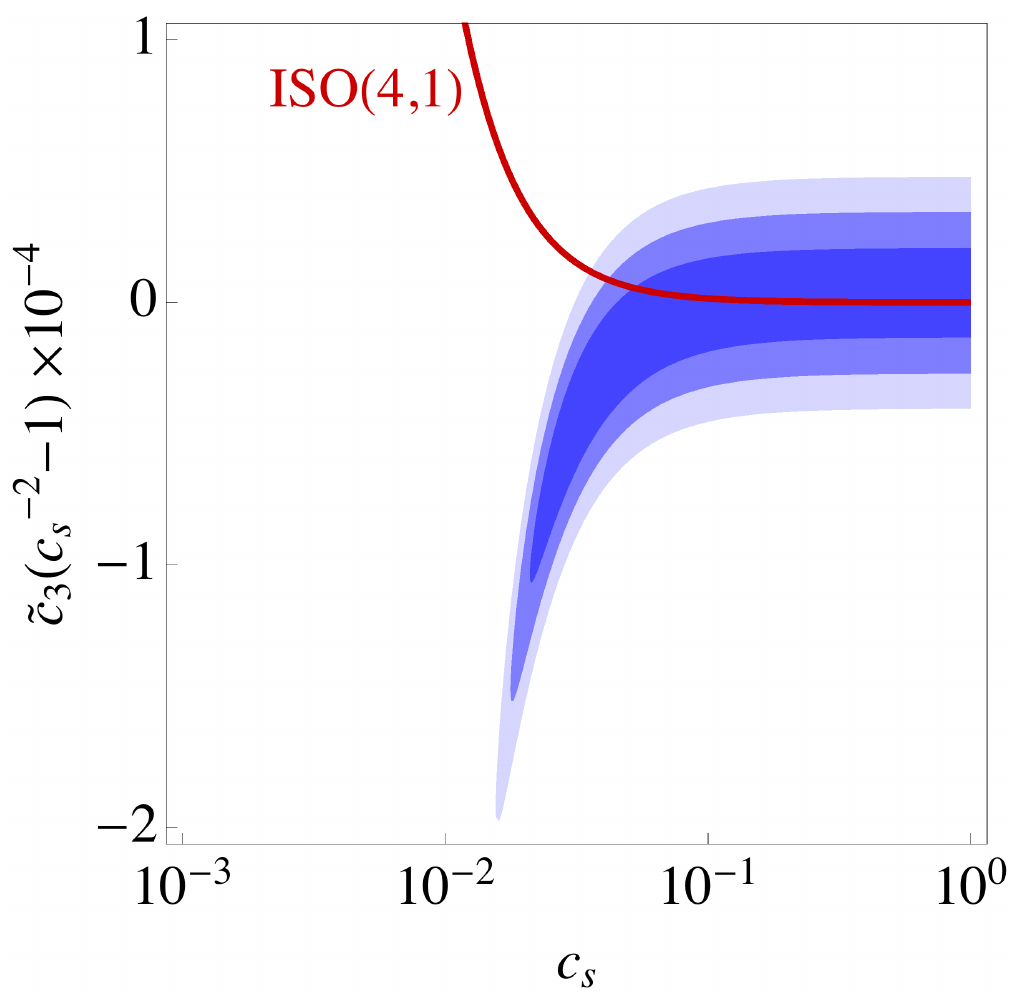}  
\includegraphics[width=0.45 \textwidth]{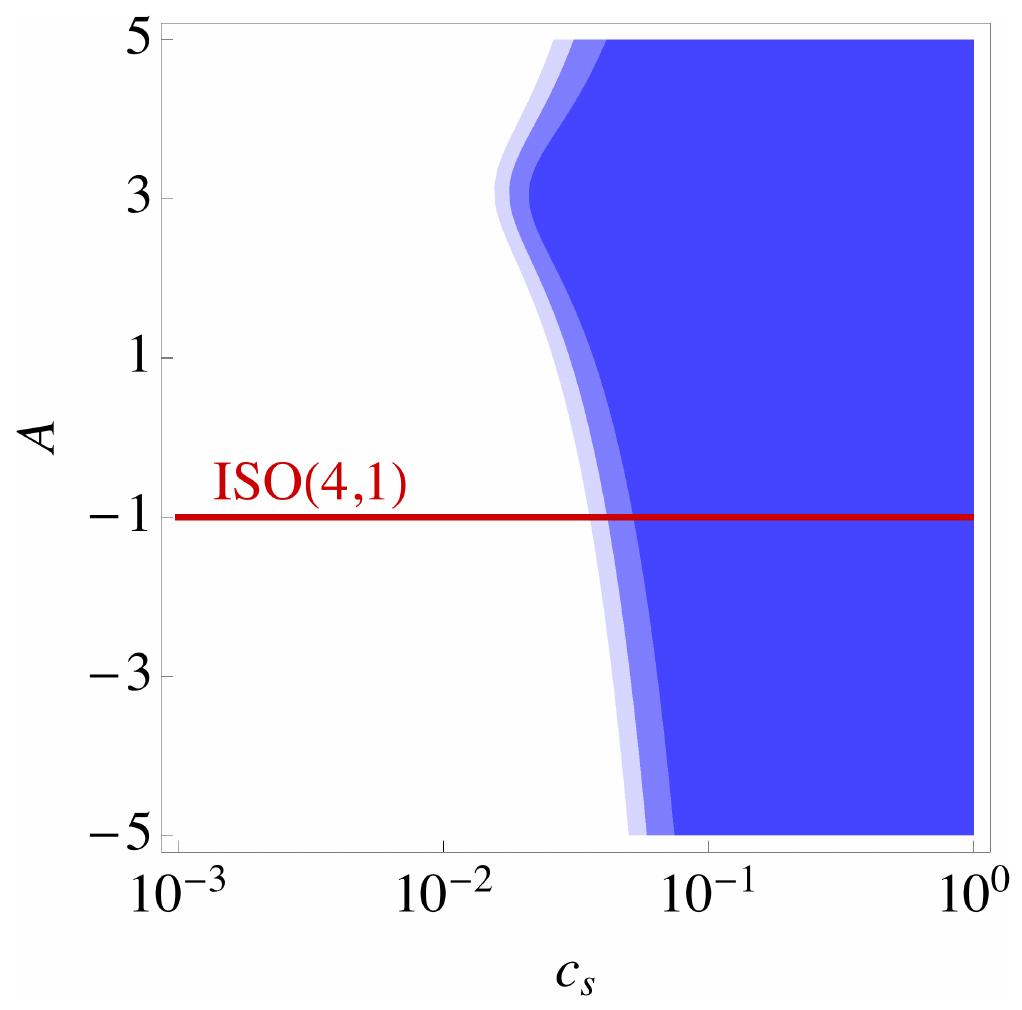}  
\end{center}
\caption{\small {\em Planck limits \cite{Ade:2013ydc}: the 68\%, 95\% and 99.7\% regions in the parameter space $(c_s,\tilde c_3)$ (above) and $(c_s, A)$ (below). The red line shows the prediction if one imposes the ISO(4,1) symmetry (the same as in DBI inflation).}}
\label{fig:Planck}
\end{figure}

As we are going to discuss later, these results will not change when gravity is taken into account. In the notation of \cite{Senatore:2009gt}
\begin{align}
S_3=\int \mathrm d^4x & \sqrt{-g} \;\dot H \mpl^2 (1-c_s^{-2}) \left[ -\frac{1}{a^2}\dot\pi(\partial_i\pi)^2 \right. \nonumber \\
& + \left. \left( 1+\frac23 \frac{\tilde c_3}{c_s^2}\right)\dot\pi^3 \right] \;,
\end{align}
the coefficient $\tilde c_3$ (that is in general free), is fixed by ISO(4,1): $\tilde c_3 = \frac32(1-c_s^2)$. In terms of the relative coefficient between the two operators $A \equiv -(c_s^2+\frac23 \tilde c_3)$, the symmetry fixes $A=-1$. The Planck limits \cite{Ade:2013ydc} on these parameters are shown in Fig.~\ref{fig:Planck}.

We can go to higher order and set to zero the cubic variation of the action \eqref{effective1}. We get a simple system of algebraic equations whose solution is 
\begin{equation}
\label{quarticrel}
\begin{split}
g_1&=a_1\frac{1-c_s^2}{c_s^4}\left(\frac54 - c_s^2 \right) \;, \\
g_2&=-a_1\frac{1-c_s^2}{c_s^2}\left(\frac32-c_s^2\right) \;, \quad g_3=a_1\frac 14 (1- c_s^2)\;.
\end{split}
\end{equation}
Again all the coefficients are completely fixed in terms of a single parameter, the speed of sound $c_s$. This does not come as a surprise: the only operator with one derivative per field, that {\em linearly} realizes the 4D Poincar\'e group and non-linearly realizes ISO(4,1) is the brane tension operator 
\begin{equation} \label{eq.action}
S=M^4 \int \mathrm d^4 x \left(1-\sqrt{1+(\partial\phi)^2}\right)\ ,
\end{equation}
so it is not surprising that everything is fixed for operators with one derivative per field. One can check that expanding \eqref{eq.action} around $\phi_0=vt$ one gets operators which satisfy \eqref{cubicrel} and \eqref{quarticrel}. Still it is nice to see the constraints directly at the level of the EFT of inflation, without assuming to be able to extrapolate far from the inflationary solution.

One can also explore the consequences of ISO(4,1) for operators with more derivatives. If we look at operators with two derivatives on one of the $\pi$'s then the effective action starts with cubic terms (quadratic terms are total derivatives) and reads 
\begin{align}
\label{effective2}
S = \int & \mathrm d^4 x \left( \lambda_1 \dot\pi^2 \partial_i^2\pi + \lambda_2 (\partial_i\pi)^2 \partial_i^2\pi \right. \nonumber \\
& \left. + \mu_1 \dot\pi^3 \partial_i^2\pi + \mu_2\dot\pi (\partial_i\pi)^2 \partial_i^2\pi  + \cdots \right)\ .
\end{align}
Using the transformation \eqref{transf1} we can easily find the relations among $\lambda_1$, $\lambda_2$, $\mu_1$ and $\mu_2$
\be
\label{highderivrel}
\lambda_2=\frac {-c_s^2}{2}\lambda_1 \;, \;\; \mu_1=\frac 43 \frac{1-c_s^2}{c_s^2} \lambda_1 \;, \;\; \mu_2= (c_s^2-1) \lambda_1\;.
\ee
As a check, one can start from the brane picture and consider an operator with one extra derivative on $\pi$ compared to the brane tension: there is only one, the extrinsic curvature of the brane. This gives the following operator which non-linearly realizes ISO(4,1) \cite{deRham:2010eu}
\be
\label{cubicgalileon}
S = M^3 \int \mathrm d^4x \;\frac{1}{1+(\partial\phi)^2} \partial_\mu\partial_\nu\phi\partial^\mu\phi\partial^\nu\phi\;.
\ee
Indeed, expanding \eqref{cubicgalileon} around $\phi_0=vt$ we find that the cubic action for the Goldstone is
\be
\label{cubicgalileonpi}
S_{3} = M^3 \int \mathrm d^4x \left(\frac{1-c_s^2}{c_s^2}\dot\pi^2\partial_i^2\pi + \partial_\mu\partial_\nu\pi\partial^\mu\pi\partial^\nu\pi \right) ,
\ee
which satisfy the constraints \eqref{highderivrel}.

\vspace{.3cm}

{\em The limit of Galilean symmetry and the coupling with gravity.}
The ISO(4,1) transformation \eqref{transf1} contains a dimensionless parameter $v$, which can be  interpreted in a 5D picture as the brane velocity in the bulk. As we discussed, this parameter fixes the speed of sound of perturbations, eq.~\eqref{cs2v}. One can consistently take the limit $v \to 0$ of the symmetry\footnote{Notice that this simply corresponds to the non-relativistic limit,  when the brane motion is slow compared to the speed of light. This does {\em not} imply that the 4D Poincar\'e symmetry is restored. Indeed the transformation of $\pi$ under a 4D boost parametrized by $\beta^i$ is given by
\be
\delta\pi = \beta^i x^i +\dot\pi \beta^i x^i +\partial_i \beta^i t \;.
\ee
This does not depend on $v$ and is still non-linearly realized for $v \to 0$.
}. 
This is a group contraction and in this limit the symmetry does not act on coordinates anymore and it thus commutes with the 4D Poincar\'e group. It reduces to an internal symmetry acting on $\pi$ only
\be
\delta \pi= \omega_\mu x^\mu  \;.
\ee
This is the Galilean symmetry studied in \cite{Nicolis:2008in}, whose implications for the EFT of inflation have been discussed in \cite{Creminelli:2010qf} (see also \cite{Burrage:2010cu}). This symmetry requires $c_s =1$ and forbids all interactions with a single derivative per field. All interactions come from higher derivative terms. For example in eq.~\eqref{cubicgalileonpi}, for $c_s = 1$ we have only the second operator which can be written as $(\partial\pi)^2 \Box\pi$, i.e.~the cubic Galileon. 

So far we discussed the ISO(4,1) symmetry in Minkowski space, without including gravity. Ultimately we are interested in calculating correlation functions during inflation, so that the coupling with gravity cannot be neglected. Similarly to what happens in the case of the Galilean symmetry discussed above, gravity breaks the ISO(4,1) symmetry\footnote{On a curved background, one cannot consistently define the constant vector $\omega^\mu$ that appears in eq.~\eqref{transf1}: this shows that the symmetry is ill-defined in the presence of gravity.}.  This implies that the symmetry is not a good one for the background evolution, since in general the Hubble friction plays an important role. This is an additional motivation to formulate the symmetry directly in the EFT of inflation as a non-linearly realized symmetry for $\pi$ on scales much shorter than Hubble, without reference to the background solution. 

Another point to address is whether the actions for $\pi$ derived above can be used, once minimally coupled to gravity, to calculate observables during inflation or gravity will completely change the picture. The breaking of the symmetry due to gravity will manifest in two ways. First of all, graviton radiative corrections will induce operators which do not respect the symmetry. This effect is arguably small, as suppressed by powers of $\mpl$. Second, in calculating $\pi$ loops on a gravitational background, non-invariant terms will also be generated. These operators will be invariant under a shift of $\pi$, as the shift symmetry is compatible with the coupling with gravity, but not fully ISO(4,1) invariant. As these terms arise only on a curved background they will contain powers of the Riemann tensor, schematically
\be
(R_{\mu\nu\rho\sigma})^n (\partial\pi)^m \;.
\ee
On a quasi de Sitter background $R \simeq H^2$, so we expect these terms to be suppressed with respect to the ones we considered above by powers of $(H/\Lambda)^2 \ll 1$, where $\Lambda$ is the UV cut-off of the theory. 

These corrections can become relevant if the coefficient of some operator is unnaturally large. For example, the effect of the induced gravity term on a brane is studied in \cite{RenauxPetel:2011uk,Renaux-Petel:2013ppa} and the conclusion is that the cubic action is in general not uniquely fixed in terms of $c_s$: a different linear combination of the operators $\dot\pi^3$ and $\dot\pi(\partial_i\pi)^2$ is possible, giving in particular an orthogonal shape of non-Gaussianity. This is at first surprising as the model respect the ISO(4,1) symmetry we are discussing. However, the deviations are indeed due to cubic operators with more than three derivatives in the EFT of inflation \cite{Renaux-Petel:2013ppa}: in curved space some of these derivatives can be traded for the curvature scale $H$ and one is left with only three derivatives on $\pi$. However a basic tenet of the EFT approach is that operators of higher dimension give small corrections: if they induce ${\cal O}(1)$ changes, it is not clear why one can neglect all the other higher dimensional terms.


\vspace{.3cm}

{\em ISO(4,1) or SO(4,2)?} In DBI inflation \cite{Alishahiha:2004eh} a probe brane lives in an AdS throat and non-linearly realizes the SO(4,2) group, so that one may wonder why we did not consider this group instead of ISO(4,1). One simple answer is that during inflation the brane does not move much in units of the AdS radius $L$, so that the difference between flat and curved bulk is immaterial. It is still interesting to understand whether SO(4,2) would give the same predictions.  

The answer is no. It is straightforward to check, for example supplementing the DBI action with other SO(4,2)-invariant operators like the AdS conformal Galileons \cite{deRham:2010eu}, that the nice predictions of ISO(4,1) are lost. In particular the speed of sound is not fixed in terms of the velocity $v$ in the bulk and the cubic operators $\dot\pi^3$ and $\dot\pi (\partial_i\pi)^2$ can appear in a general linear combination. The fact that $c_s^2$ is not fixed in terms of $v$ may come as a surprise: after all it simply comes from the relativistic sum of velocities and this should apply locally also in AdS. This intuition however requires that higher derivative operators are suppressed by a cutoff scale $\Lambda \gg L^{-1}$: in this case only the tension of the brane is important and we get back to the DBI inflation case. When, on the other hand, $\Lambda \sim L^{-1}$ higher derivative operators are unsuppressed, the brane is a thick object in comparison with the AdS radius: it will not follow geodesics and we do not expect the same predictions as for DBI inflation, though the SO(4,2) symmetry is preserved.

All this can also be seen at the level of the EFT. The most general action allowed by the symmetry up to quadratic order is 
\be
S_\text{EFT}=\int \mathrm d^4 x \, \left( a_0 \pi +a_1 \dot{\pi}^2 +a_2(\partial_i\pi)^2+ m^2 \pi^2 \right)\;,
\ee
where all the coefficients are now time dependent. As in the ISO(4,1) case, a background solution with constant velocity is not in general a solution, therefore we have to keep $a_0$ that will be cancelled by additional terms which are not SO(4,2) symmetric. The  non-linear transformation of $\pi$ that realizes SO(4,2) is
\be\label{eq.so42transf}
\begin{split}
\delta \pi = \frac{1}{\dot \phi_0} \bigg( &\omega_\mu x^\mu \phi+\omega_\mu x^\mu x^\nu\partial_\nu \phi -\frac 12 x^2 \omega^\mu \partial_\mu \phi\\
&+\frac 12 \omega^\mu \partial_\mu \phi-\frac{1}{2\phi^2} \omega^\mu \partial_\mu \phi  \bigg)\;.
\end{split}
\ee
Again, requiring the invariance of the action under this transformation leads to a set of three constraints on the coefficients 
\begin{equation}\label{eq.constrSO42}
\begin{split}
&m^2-\frac{1}{2}\dot a_0+\frac{\ddot \phi}{2\dot \phi}a_0=0 \ ,\\
&3 a_0+4 \dot a_1-2\partial_t \left( a_1 \frac{\ddot \phi_0 \phi_0}{\dot \phi_0^2}\right)-2 \frac{\phi_0}{\dot\phi_0} m^2 =0 \ ,\\
&6 a_2+4 a_1+2\partial_t \left( a_1 \frac{\phi_0^4-\dot \phi_0^2}{ \phi_0^3\dot \phi_0}\right)-2\frac{\ddot \phi_0 \phi_0}{\dot \phi_0^2}a_1=0\  .
\end{split}
\end{equation}
It is straightforward to verify that, for a constant $\dot\phi_0$, these constraints do not fix the relation among the coefficients of the kinetic term and therefore $c_s$. 

A particular case in which these constraints actually fix the form of the speed of sound, is the one in which the background preserves an SO(4,1) subgroup of SO(4,2) \cite{Hinterbichler:2012fr}.
This happens for a background solution $\phi_0(t)= \alpha \;t^{-1}$.   In this case the tadpole term is absent and $\pi$ is massless so that the constraints in \eqref{eq.constrSO42} greatly simplify. The speed of sound is then fixed by the last constraint and the residual dilation symmetry \cite{Hinterbichler:2012fr}: 
$c_s^2 = 1- \alpha^{-2}$. 

\vspace{.3cm}

{\em Conclusions}. Imposing an ISO(4,1) symmetry in the EFT of inflation leaves (at leading order in derivative) a single free parameter, the speed of sound $c_s$. It should be straightforward to study the consequences of ISO(4,1) in the EFT of multi-field inflation \cite{Senatore:2010wk}: this symmetry is indeed at play in multi-field DBI models 
\cite{Langlois:2008wt}. 

\vspace{.3cm}

\noindent
{\em Acknowledgements.}
It is a pleasure to thank M.~Serone for useful discussions, N.~Bartolo and M.~Liguori for correspondence about Planck data.


\begin{thebibliography}{99}

\bibitem{Cheung:2007st} 
  C.~Cheung, P.~Creminelli, A.~L.~Fitzpatrick, J.~Kaplan, and L.~Senatore,
  ``The Effective Field Theory of Inflation,''
  JHEP {\bf 0803}, 014 (2008)
  [arXiv:0709.0293 [hep-th]].
  
\bibitem{Ade:2013ydc} 
  P.~A.~R.~Ade {\it et al.}  [Planck Collaboration],
  ``Planck 2013 Results. XXIV. Constraints on primordial non-Gaussianity,''
  arXiv:1303.5084 [astro-ph.CO].
  
\bibitem{Alishahiha:2004eh} 
  M.~Alishahiha, E.~Silverstein, D.~Tong and ,
  ``DBI in the sky,''
  Phys.\ Rev.\ D {\bf 70}, 123505 (2004)
  [hep-th/0404084].
 
\bibitem{Ade:2013uln} 
  P.~A.~R.~Ade {\it et al.}  [Planck Collaboration],
  ``Planck 2013 results. XXII. Constraints on inflation,''
  arXiv:1303.5082 [astro-ph.CO].
  
\bibitem{Senatore:2009gt} 
  L.~Senatore, K.~M.~Smith and M.~Zaldarriaga,
  ``Non-Gaussianities in Single Field Inflation and their Optimal Limits from the WMAP 5-year Data,''
  JCAP {\bf 1001}, 028 (2010)
  [arXiv:0905.3746 [astro-ph.CO]].
  
\bibitem{deRham:2010eu} 
  C.~de Rham and A.~J.~Tolley,
  ``DBI and the Galileon reunited,''
  JCAP {\bf 1005}, 015 (2010)
  [arXiv:1003.5917 [hep-th]].
  
\bibitem{Nicolis:2008in} 
  A.~Nicolis, R.~Rattazzi and E.~Trincherini,
  ``The Galileon as a local modification of gravity,''
  Phys.\ Rev.\ D {\bf 79}, 064036 (2009)
  [arXiv:0811.2197 [hep-th]].
  
\bibitem{Creminelli:2010qf} 
  P.~Creminelli, G.~D'Amico, M.~Musso, J.~Norena and E.~Trincherini,
  ``Galilean symmetry in the effective theory of inflation: new shapes of non-Gaussianity,''
  JCAP {\bf 1102}, 006 (2011)
  [arXiv:1011.3004 [hep-th]].
    
\bibitem{Burrage:2010cu} 
  C.~Burrage, C.~de Rham, D.~Seery and A.~J.~Tolley,
  ``Galileon inflation,''
  JCAP {\bf 1101}, 014 (2011)
  [arXiv:1009.2497 [hep-th]].

\bibitem{RenauxPetel:2011uk} 
  S.~Renaux-Petel, S.~Mizuno and K.~Koyama,
  ``Primordial fluctuations and non-Gaussianities from multifield DBI Galileon inflation,''
  JCAP {\bf 1111}, 042 (2011)
  [arXiv:1108.0305 [astro-ph.CO]].
  
\bibitem{Renaux-Petel:2013ppa} 
  Séb.~Renaux-Petel,
  ``DBI Galileon in the Effective Field Theory of Inflation: Orthogonal non-Gaussianities and constraints from the Trispectrum,''
  arXiv:1303.2618 [astro-ph.CO].
  
\bibitem{Hinterbichler:2012fr} 
  K.~Hinterbichler, A.~Joyce, J.~Khoury and G.~E.~J.~Miller,
  ``DBI Realizations of the Pseudo-Conformal Universe and Galilean Genesis Scenarios,''
  JCAP {\bf 1212}, 030 (2012)
  [arXiv:1209.5742 [hep-th]].
  
\bibitem{Senatore:2010wk} 
  L.~Senatore and M.~Zaldarriaga,
  ``The Effective Field Theory of Multifield Inflation,''
  JHEP {\bf 1204}, 024 (2012)
  [arXiv:1009.2093 [hep-th]].
  
\bibitem{Langlois:2008wt} 
  D.~Langlois, S.~Renaux-Petel, D.~A.~Steer and T.~Tanaka,
  ``Primordial fluctuations and non-Gaussianities in multi-field DBI inflation,''
  Phys.\ Rev.\ Lett.\  {\bf 101}, 061301 (2008)
  [arXiv:0804.3139 [hep-th]].
   
\end{thebibliography}
\end{document}